\documentclass[11pt]{article}
\usepackage{amsmath,amsthm,amscd,array,amssymb}

\begin{document}

\begin{center}
{\Large{\bf Twisted $N = 8$, $D = 2$ super Yang--Mills theory as 
\\
\medskip\smallskip
 example of a Hodge--type cohomological theory }}
\\
\bigskip\medskip
{\large{\sc B. Geyer}}$^a$
\footnote{Email: geyer@itp.uni-leipzig.de}
and 
{\large{\sc D. M\"ulsch}}$^{b}$
\footnote{Email: muelsch@informatik.uni-leipzig.de}
\\
\smallskip
{\it $^a$ Universit\"at Leipzig, Naturwissenschaftlich-Theoretisches Zentrum
\\
$~$ and Institut f\"ur Theoretische Physik, D--04109 Leipzig, Germany
\\\smallskip
$\!\!\!\!\!^b$ Wissenschaftszentrum Leipzig e.V., D--04103 Leipzig, Germany}
\\
\bigskip
{\small{\bf Abstract}}
\\
\end{center}

\begin{quotation}
\noindent {\small{It is shown that the dimensional reduction of the 
$N_T = 2$, $D = 3$ Blau--Thompson model to $D = 2$, i.e., the novel 
topological twist of $N = 8$, $D = 2$ super--Yang--Mills theory, 
provides an example of a Hodge--type cohomological theory. In that
theory the generators of the topological shift, co--shift and gauge 
symmetry, ($Q^a$, $^\star Q^a$, $G$), together with a discrete duality 
operation are completely analogous to the de Rham cohomology operators 
($d, \delta, \Delta$) and the Hodge $\star$--operation.}}
\end{quotation}

\bigskip\medskip
%%%%%%%%%%%%%%%%%%%%%%%%%%%%%%%%%%%%%%%%%%%%%%%%%%%%%%%%%%%%%%%%%%%%%%%%%%%%%
\begin{flushleft}
{\large{\bf 1. Introduction}}
\end{flushleft}
%%%%%%%%%%%%%%%%%%%%%%%%%%%%%%%%%%%%%%%%%%%%%%%%%%%%%%%%%%%%%%%%%%%%%%%%%%%%%
\bigskip
Some very enlightening attemps \cite{1} -- \cite{8} have been made to 
incorporate into the gauge--fixing procedure of general gauge theories besides 
the basic ingredience of BRST cohomology $Q$ also a co--BRST cohomology 
$^\star Q$ which, together with the BRST Laplacian $\Delta$, form the 
same kind of superalgebra as their counterparts, the de Rham cohomology 
operators $d$, $\delta = \pm \star d \star$ and $\Delta = d \delta + \delta d$, 
in differential geometry \cite{9}, namely,
\begin{alignat*}{3}
Q^2 &= 0, 
&\qquad  
^\star Q^2 &= 0,
&\qquad 
^\star Q &= \pm \star Q \star,
\\
[ \Delta, Q ] &= 0,
&\qquad 
[ \Delta, \,^\star Q ] &= 0,
&\qquad
\Delta &= \{ Q, \,^\star Q \} \neq 0.
\end{alignat*}
In Ref. \cite{8} it has been proven that whenever the ghost--extended 
quantum state space $H_{\rm ext}$ possesses a non--degenerate inner product,
which is also non--degenerated upon restriction to the image of $Q$,Im\, $Q$, it allows the 
following complete and orthogonal Hodge decomposition:
\begin{equation*}  
H_{\rm ext} = {\rm Ker}\, \Delta + {\rm Im}\, Q + {\rm Im} \,^\star Q.
\end{equation*}
In terms of quantum state vectors, this decomposition theorem asserts 
that an arbitrary state $\psi \in H_{\rm ext}$ can be uniquely decomposed 
into a sum of a BRST exact, a co--BRST exact and a BRST harmonic state,
\begin{equation*}
\psi = \omega + Q \chi + \,^\star Q \phi,
\qquad
\Delta \omega = 0
\qquad
\Leftrightarrow
\qquad
Q \omega = 0,
\quad
^\star Q \omega = 0.
\end{equation*}
The physical properties of $\psi$ lie entirely within the 
BRST harmonic state $\omega$ which is given by the zero modes of the
operator $\Delta$. In proving the consistency of the BRST quantization
procedure the essential difficulty consists in showing that the BRST
cohomology defines physical (BRST singlet) states 
$\psi_{\rm phys} \in H_{\rm phys} = {\rm Ker}\, Q/ {\rm Im}\, Q$, which are
taken to be BRST--closed but not exact. Usually, in order to select from 
$H_{\rm ext}$ the physical state space $H_{\rm phys}$, one imposes the 
BRST gauge condition $Q \psi = 0$ \cite{10} -- \cite{12}. However,
on the class of BRST--closed states $\psi = \omega + Q \chi$, which satisfy 
this condition, besides the harmonic states $\omega$ there exists also 
a set of spurios BRST--exact states, $Q \chi$, which have zero physical
norm. On the other hand, by imposing the co--BRST gauge condition, 
$^\star Q \psi = 0$, as well one gets for each BRST cohomology class the 
uniquely determined harmonic states, $\psi = \omega$.

It has been a long--standing problem to give a prototype example of the
BRST gauge--fixing procedure based on harmonic gauges for a field theoretical
model. Recently, it has been shown that the $D = 2$ Maxwell theory can 
be formulated as being a topological field theory of Witten type with 
underlying symmetries being 
of Schwarz type \cite{13}. This topological theory provides an example 
of a Hodge--type cohomological theory where, not only all the de Rham 
cohomology operators ($d$, $\delta$, $\Delta$) are expressed in terms of 
generators of some local symmetries, but even an analogue of the Hodge 
duality ($\star$) operation exists as discrete 
symmetry of that theory. On the other hand, pure 
$D = 2$ Yang--Mills theory does not such a nice structure, because in that 
theory there is no discrete symmetry inter--relating the BRST and co--BRST 
charges by some duality operation \cite{14}.    

In this paper we give another example for a Hodge--type cohomological theory, 
but this time for a non--abelian gauge theory. More precisely, we consider 
a Witten type topological theory in $D = 2$ with an underlying $N_T = 4$ 
equivariant cohomology. This theory is obtained either by a topological 
twist of $N = 8$, $D = 2$ super--Yang--Mills theory (SYM) \cite{15} or by a 
dimensional reduction to $D = 2$ of the $N_T = 2$ novel topological twist of 
$N = 4$, $D = 3$ SYM constructed by Blau and Thompson \cite{16}. The four 
nilpotent topological supercharges of that theory, 
denoted by $Q^a = (Q, \bar{Q})$ and 
$^\star Q^a = (^\star Q, \,^\star \bar{Q})$, $a = 1,2$, together with the 
generator $G$ of the gauge transformations, obey the same 
kind of superalgebra as their counterparts in differential geometry,
the de Rham cohomology operators $d$, $\delta = \pm \star d\, \star$ 
and $\Delta$. More precisely, there exists an direct analogy between 
the two sets of cohomology operators and the two sets of symmetry operators 
of the topological theory according to the following assignments:  
\begin{alignat}{4}
\label{2.1}
d 
&\quad &\Leftrightarrow \quad&
Q
&\qquad\qquad 
d 
&\quad &\Leftrightarrow \quad&
\bar{Q}
\nonumber
\\
\delta = \star\, d\, \star 
&\quad &\Leftrightarrow \quad&
\,^\star \bar{Q} = \star\, Q \star
&\qquad\qquad
\delta = - \star d\, \star 
&\quad &\Leftrightarrow \quad&
\,^\star Q = - \star \bar{Q} \star
\\
\Delta = \{ d, \delta \} 
&\quad &\Leftrightarrow \quad&
\{ Q, \,^\star \bar{Q} \} = G
&\qquad\qquad
\Delta = \{ d, \delta \} 
&\quad &\Leftrightarrow \quad&
\{ \bar{Q}, \,^\star Q \} = - G.
\nonumber
\end{alignat}
That is, the exterior and the co--exterior derivatives, $d$ and $\delta$,
are related to the nilpotent topological shift and co--shift operators
($Q$, $\bar{Q}$) and ($\,^\star Q$, $\,^\star \bar{Q}$), respectively.
 Furthermore, it is shown that the analogue of the Laplacian $\Delta$  
is the gauge generator $G$. This is due to the fact that for the 
$N_T = 4$ equivariant cohomology the anti--commutation relations of the 
topological superalgebra, $\{ Q, \,^\star \bar{Q} \} = G$ and 
$\{ \bar{Q}, \,^\star Q \} = - G$, close only modulo the gauge transformation 
$G =2 \delta_G(\phi)$ generated by a $Q^a$-- and $^\star Q^a$--invariant 
scalar field $\phi$. 
Moreover, it will be shown that the topological supercharges are 
interrelated by an exact discrete symmetry of that theory which may be
interpreted as the corresponding duality ($\star$) operation. 

The outline of the paper is as follows. In Section 1 we construct
$N_T = 4$, $D = 2$ topological Yang--Mills theory (TYM) by performing, first, 
a dimensional reduction of $N = 1$, $D = 6$ SYM to $D = 2$ and, afterwards,
a topological twist analogous to the one used in constructing the  
$N_T = 2$ Blau--Thompson model in $D = 3$ \cite{16}.
In Section 2 we give a realization of the de Rham cohomology operators in 
terms of generators of the topological shift and co--shift symmetries. 
Furthermore, in Section 3, we show that they are related to each other by a 
discrete 
symmetry of the theory having the property of a Hodge--type duality operation. 
In addition, we give the transformation rules for the vector and co--vector 
supersymmetries.
\bigskip
%%%%%%%%%%%%%%%%%%%%%%%%%%%%%%%%%%%%%%%%%%%%%%%%%%%%%%%%%%%%%%%%%%%%%%%%%%%%%
\begin{flushleft}
{\large{\bf 2. The topological twist of $N = 8$, $D = 2$ 
super--Yang--Mills theory}}
\end{flushleft}
%%%%%%%%%%%%%%%%%%%%%%%%%%%%%%%%%%%%%%%%%%%%%%%%%%%%%%%%%%%%%%%%%%%%%%%%%%%%%
\bigskip
In this section we construct the $N_T = 4$, $D = 2$ topological gauge theory 
with an underlying equivariant cohomology whose four supercharges
$Q^a$ and $^\star Q^a$, $a=1,2$, together with the gauge generator $G$ 
obey the cohomological algebra (\ref{2.1}). In order to get that theory
we dimensional reduce and twist the $N = 1$, $D = 6$ SYM 
in the Euclidean space--time. Since the details of this procedure are 
straightforward, we will only focus on the relevant steps.

The Euclidean action of $N = 1$, $D = 6$ SYM,
\begin{equation}
\label{2.2}
S^{(N = 1)} = \int d^6x\, {\rm tr} \Bigr\{
\hbox{\large$\frac{1}{4}$} F^{MN} F_{MN} +
i \bar{\lambda} \Gamma^M D_M \lambda \Bigr\},
\end{equation}
is build up from an anti--hermitean vector field $A_M$ ($M = 1, \ldots 6$) 
and a complex chiral spinor $\lambda$ in the adjoint representation of the 
gauge group. This action is invariant under the following supersymmetry
transformations:
\begin{gather}
\label{2.3}
\delta_Q A_M = i \bar{\lambda} \Gamma_M \epsilon - 
i \bar{\epsilon} \Gamma_M \lambda,
\qquad
%\nonumber \\
\delta_Q \lambda = - \hbox{\large$\frac{1}{2}$} 
\Gamma^M \Gamma^N F_{MN} \epsilon,
\qquad
\delta_Q \bar{\lambda} = \hbox{\large$\frac{1}{2}$}
\bar{\epsilon} \Gamma^M \Gamma^N F_{MN},
\end{gather}
$\epsilon$ being a chiral spinor. For the $8$--dimensional Euclidean Dirac 
matrices, $\{ \Gamma_M, \Gamma_N \} = - 2 \delta_{MN} {\rm I}_8$, we choose 
the representation
\begin{align*}
&\Gamma_m = - i \begin{pmatrix} 0 & 
\delta_\alpha^{~\beta} (\sigma_m)_A^{\!~~B} \\
\delta_\alpha^{~\beta} (\sigma_m)_A^{\!~~B} & 0 \end{pmatrix},
\qquad
m = 1,2,3, \quad \alpha = 1,2, \quad A = 1,2,
\\
&\Gamma_{3 + m} = \begin{pmatrix} 0 & 
- (\sigma_m)_\alpha^{~\beta} \delta_A^{\!~~B} \\
(\sigma_m)_\alpha^{~\beta} \delta_A^{\!~~B} & 0 \end{pmatrix},
\qquad
\Gamma_7 = - i \begin{pmatrix} 
\delta_\alpha^{~\beta} \delta_A^{\!~~B} & 0 \\ 0 &  
- \delta_\alpha^{~\beta} \delta_A^{\!~~B} \end{pmatrix},
\end{align*}
where $\sigma_m$ are the Pauli matrices. At the first sight this 
representation seems not to be well adapted to a $4 + 2$ decomposition
of the Euclidean space--time since it does not directly expose the 
2--dimensional Dirac matrices. But, let us shortly explain why nevertheless
this representation is well suited for our purpose. After performing a
dimensional reduction of $N = 1$, $D = 6$ SYM to $D = 3$ the reduced action 
of $N = 4$, $D = 3$ SYM has a global symmetry group 
$SU(2)_E \otimes SU(2)_N \otimes SU(2)_R$ \cite{17}, $SU(2)_N$ being the 
internal Euclidean symmetry group arising from the decomposition 
$SO(6) \rightarrow SU(2)_N \otimes SU(2)_E$. The symmetry group $SU(2)_E$ 
is the Euclidean group in $D = 3$ and $SU(2)_R$ is the symmetry group
of $N = 1$, $D = 6$ SYM. There are two essentially different 
possibilities to construct topological models with $N_T = 2$ topological
supercharges, arising from twisting $N = 4$, $D = 3$ SYM. The standard twist 
consists in replacing $SU(2)_E \otimes SU(2)_R$ through is diagonal 
subgroup and gives the $N_T = 2$, $D = 3$ super--BF model \cite{18}. The 
second twist, which is intrinsically 3--dimensional, is obtained by taking 
the diagonal of $SU(2)_E \otimes SU(2)_N$ and leads to the $N_T = 2$, $D = 3$
novel topological twist introduced by Blau and Thompson \cite{16}.
This model, which is the one of interest here, has certain unusual features, 
e.g., there are no bosonic scalar fields and hence there is no 
underlying equivariant cohomology. 

After dimensional reduction $A_M$ 
becomes the $D = 3$ dimensional complexified gauge field $A_\mu \pm i V_\mu$. 
The rather striking feature of this model is the strictly nilpotency of 
the twisted supercharges $Q^a = (Q, \bar{Q})$ even prior to the introduction
of the gauge ghosts. Of course, it is not possible to identify $Q$ and
$\bar{Q}$ with the exterior and the co--exterior derivative, $d$ and
$\delta$, since, by virtue of the relative nilpotency $\{ Q, \bar{Q} \} = 0$,
the gauge generator $G$ cannot be identified with the Laplacian $\Delta$. 
But, by dimensional reducing the $N_T = 2$ novel topological model further 
to $D = 2$ the third components of $A_\mu \pm i V_\mu$ become scalar fields, 
$A_3 + i V_3 = \bar{\phi}$ and $A_3 - i V_3 = \phi$, and, therefore, 
we reach a topological gauge theory with an underlying $N_T = 4$ equivariant 
cohomology. The twisted supercharges $Q^a = (Q, \bar{Q})$ and 
$^\star Q^a = (^\star Q, \,^\star \bar{Q})$ of this theory are 
nilpotent as before but the anticommutators 
$\{ Q, \,^\star \bar{Q} \} = G$ and $\{ \bar{Q}, \,^\star Q \} = - G$ 
where the gauge transformations $G =2 \delta_G(\phi)$ are
generated by $\phi$. Moreover, this theory turns out to be also invariant 
under a discrete symmetry inter--relating both, 
($Q, \,^\star \bar{Q}$) and ($\bar{Q}, \,^\star Q$), by a 
duality operation according to (\ref{2.1}), so that we have a perfect 
example of a Hodge--type cohomological theory. 

The chiral spinor $\lambda = i \gamma_7 \lambda$ can be written as 
\begin{equation*}
\lambda = \begin{pmatrix} \lambda_{\alpha A} \\ 0 \end{pmatrix}, 
\qquad
\bar{\lambda} = ( 0, \bar{\lambda}^{\alpha A} ),
\qquad
\alpha = 1,2, \quad A = 1,2,
\end{equation*}
where the unconstrained, complex 4--spinors $\lambda_{\alpha A}$ and
$\bar{\lambda}^{\alpha A}$ transform in the fundamental and their conjugate
representation of $SU(4)$, respectively. The spinor indices $A$ are raised
and lowered as follows, $\lambda^A = \epsilon^{AB} \lambda_B$ and
$\lambda_B = \lambda^A \epsilon_{AB}$, with 
$\epsilon^{AC} \epsilon_{BC} = \delta^A_{\!~~B}$ (and analogous for $\alpha$).

We further define
\begin{equation*}
A_M = \Bigr\{
A_\mu,~ A_3 = \hbox{\large$\frac{1}{2}$} ( \phi + \bar{\phi} ),~
A_{\mu + 3} = V_\mu,~
A_6 = \hbox{\large$\frac{1}{2}$} i ( \phi - \bar{\phi} ) \Bigr\},
\qquad
\mu = 1,2.
\end{equation*}
As a next step, we suppose that no field depends on $x^3, \cdots, x^6$ and
decompose the action (\ref{2.2}) under the assumption of trivial 
dimensional reduction. Then, we obtain the reduced action of $N = 8$, $D = 2$
SYM, 
\begin{align}
\label{2.4}
S^{(N = 8)} = \int d^2x\, {\rm tr} \Bigr\{&
\hbox{\large$\frac{1}{4}$} F^{\mu\nu}(A - i V) F_{\mu\nu}(A + i V) + 
\hbox{\large$\frac{1}{2}$} D^\mu(A) V_\mu D^\nu(A) V_\nu
\nonumber
\\
& + \hbox{\large$\frac{1}{2}$} D^\mu(A) \phi D_\mu(A) \bar{\phi} +
\hbox{\large$\frac{1}{2}$} [ V^\mu, \bar{\phi} ] [ V_\mu, \phi ] - 
\hbox{\large$\frac{1}{8}$} [ \phi, \bar{\phi} ] [ \phi, \bar{\phi} ]
\nonumber
\\
& - \bar{\lambda}^{\alpha C} (\sigma^\mu)_C^{\!~~D} 
D_\mu(A) \lambda_{\alpha D} - 
i \bar{\lambda}^{\alpha C} (\sigma^\mu)_\alpha^{~\beta} 
[ V_\mu, \lambda_{\beta C} ] 
\phantom{\frac{1}{2}} 
\nonumber
\\
& - \hbox{\large$\frac{1}{2}$} \bar{\lambda}^{\alpha C}  
(\sigma_3)_C^{\!~~D} [ \phi + \bar{\phi}, \lambda_{\alpha D} ] +
\hbox{\large$\frac{1}{2}$} \bar{\lambda}^{\alpha C} 
(\sigma_3)_\alpha^{~\beta} [ \phi - \bar{\phi}, \lambda_{\beta C} ] \Bigr\},
\end{align}
where $F_{\mu\nu}(A) = \partial_{[\mu} A_{\nu]} + [ A_\mu, A_\nu ]$ is the
YM field strenght and $D_\mu(A) = \partial_\mu + [ A_\mu, ~\cdot~ ]$
the covariant derivative of the gauge field $A_\mu$. Recalling that in the 
$D = 2$ dimensional Euclidean space--time there are no propagating
degrees of freedom associated with $A_\mu$. 

In order to get from (\ref{2.4}) a topological theory we perform the same
twist as in Ref. \cite{1}, i.e., we identify the spinor index $A$ with 
$\alpha$, and decompose the twisted spinor fields as follows,
\begin{align*}
\lambda_{AB} &= \hbox{\large$\frac{1}{\sqrt{2}}$} \Bigr\{ 
\epsilon_{AB} \eta + 
(\sigma^\mu)_{AB} \psi_\mu + (\sigma_3)_{AB} \zeta \Bigr\},
\\
\bar{\lambda}^{AB} &= \hbox{\large$\frac{1}{\sqrt{2}}$} \Bigr\{
\epsilon^{AB} \bar{\eta} + 
(\sigma^\mu)_{AB} \bar{\psi}_\mu + (\sigma_3)^{AB} \bar{\zeta} \Bigr\};
\end{align*}
here, $\eta^a = (\eta, \bar{\eta})$, 
$\zeta^a = (\zeta, \bar{\zeta})$ and
$\psi_\mu^a = (\psi_\mu, \bar{\psi}_\mu)$ form $SU(2)_R$ doublets of
Grassmann--odd scalar fields and ghost--for--antighost vector fields,
respectively, $SU(2)_R$ being the internal symmetry group of the twisted 
action. Then, by making use of the following equalities, 
\begin{align*}
(\sigma_\mu)_A^{\!~~C} (\sigma_\nu)_{CB} &= 
\delta_{\mu\nu} \epsilon_{AB} + i \epsilon_{\mu\nu} (\sigma_3)_{AB},
\qquad
\mu = 1,2,
\\
(\sigma_\mu)_A^{\!~~C} (\sigma_3)_{CB} &= 
- i \epsilon_{\mu\nu} (\sigma^\nu)_{AB},
\qquad
(\sigma_3)_A^{\!~~C} (\sigma_3)_{CB} = \epsilon_{AB},
\end{align*}
where $\epsilon_{\mu\nu}$, $\epsilon^{\mu\rho} \epsilon_{\nu\rho} = 
\delta^\mu_{~\nu}$, is the anti--symmetric Levi--Civita tensor in $D = 2$, 
from (\ref{2.4}) we arrive at the twisted action of $N_T = 4$, $D = 2$ TYM 
we are looking for,
\begin{align}
\label{2.5}
S_T^{(N_T = 4)} = \int d^2x\, {\rm tr} \Bigr\{&
\hbox{\large$\frac{1}{4}$} F^{\mu\nu}(A - i V) F_{\mu\nu}(A + i V) + 
\hbox{\large$\frac{1}{2}$} (D^\mu(A) V_\mu \; D^\nu(A) V_\nu
\nonumber
\\
& - i \epsilon^{\mu\nu} \epsilon_{ab} \zeta^a D_\mu(A + i V) \psi_\nu^b - 
\epsilon_{ab} \eta^a D^\mu(A - i V) \psi_\mu^b
\nonumber
\\
& + \hbox{\large$\frac{1}{2}$} D^\mu(A - i V) \phi\; D_\mu(A + i V) \bar{\phi} -
\hbox{\large$\frac{1}{2}$} i [ \phi, \bar{\phi} ] D^\mu(A) V_\mu
\phantom{\frac{1}{2}}
\nonumber
\\
& - \hbox{\large$\frac{1}{8}$} [ \phi, \bar{\phi} ] [ \phi, \bar{\phi} ] -
\hbox{\large$\frac{1}{2}$} i \epsilon^{\mu\nu} 
\epsilon_{ab} \bar{\phi} \{ \psi_\mu^a, \psi_\nu^b \} +
\epsilon_{ab} \phi \{ \eta^a, \zeta^b \} \Bigr\}.
\end{align}
The internal indices $a$ are raised and lowered as follows,
$\varphi^a = \epsilon^{ab} \varphi_b$ and 
$\varphi_b = \varphi^a \epsilon_{ab}$, with $\epsilon_{ab}$, 
$\epsilon^{ac} \epsilon_{bc} = \delta^a_{~b}$, being the invariant tensor of 
$SU(2)_R$.
\bigskip
%%%%%%%%%%%%%%%%%%%%%%%%%%%%%%%%%%%%%%%%%%%%%%%%%%%%%%%%%%%%%%%%%%%%%%%%%%%%%
\begin{flushleft}
{\large{\bf 3. Realization of de Rham cohomology operators and 
Hodge $\star$--operation }}
\end{flushleft}
%%%%%%%%%%%%%%%%%%%%%%%%%%%%%%%%%%%%%%%%%%%%%%%%%%%%%%%%%%%%%%%%%%%%%%%%%%%%%
\bigskip
We now want to show that the topological gauge theory constructed above is 
indeed an example of a Hodge--type cohomological theory. To begin with, 
it is convenient to cast the action 
(\ref{2.5}) in the form
\begin{align}
\label{3.1}
S_T = \hbox{\large$\frac{1}{2}$} \int d^2x\, {\rm tr} \Bigr\{&
i \epsilon^{\mu\nu} \bar{B} F_{\mu\nu}(A - i V) - 
i \epsilon^{\mu\nu} B F_{\mu\nu}(A + i V) - 4 B \bar{B}
\nonumber
\\ 
& - 2 i \epsilon^{\mu\nu} \epsilon_{ab} \zeta^a D_\mu(A + i V) \psi_\nu^b - 
2 \epsilon_{ab} \eta^a D^\mu(A - i V) \psi_\mu^b 
\nonumber
\\
& + D^\mu(A) \phi D_\mu(A) \bar{\phi} + 
[ V^\mu, \phi ] [ V_\mu, \bar{\phi} ] - 
4 Y D^\mu(A) V_\mu - 4 Y^2
\phantom{\frac{1}{2}}
\nonumber
\\
& - \hbox{\large$\frac{1}{4}$} [ \phi, \bar{\phi} ] [ \phi, \bar{\phi} ] -
i \epsilon^{\mu\nu} \epsilon_{ab} \bar{\phi} \{ \psi_\mu^a, \psi_\nu^b \} +
2 \epsilon_{ab} \phi \{ \eta^a, \zeta^b \} - 
4 E^\mu \bar{E}_\mu \Bigr\},
\end{align}
where we have introduced a set of auxiliary fields, namely the Grassmann--even
scalar fields $B$, $\bar{B}$, $Y$ and the Grassmann--even vector fields
$E_\mu$, $\bar{E}_\mu$.
 
Then, the action (\ref{3.1}) can be rewritten as a sum of a topological 
BF--like term and a $Q^a$--exact term, which bears a close resemblance
to the $N_T = 2$ Blau--Thompson model in $D = 3$ \cite{6}, 
\begin{equation*}
S_T = \hbox{\large$\frac{1}{2}$} \int d^2x\, {\rm tr} \Bigr\{
i \epsilon^{\mu\nu} \bar{B} F_{\mu\nu}(A - i V) \Bigr\} + 
\hbox{\large$\frac{1}{2}$} \epsilon_{ab} Q^a Q^b X,
\qquad
Q^a = \begin{pmatrix} Q \\ \bar{Q} \end{pmatrix},
\end{equation*}
with the gauge boson
\begin{equation*}
X = - \int d^2x\, {\rm tr} \Bigr\{
\hbox{\large$\frac{1}{2}$} \bar{\phi} \bigr(
\bar{B} + \hbox{\large$\frac{1}{4}$} i \epsilon^{\mu\nu} 
F_{\mu\nu}(A + i V) \bigr) + i \bar{E}^\mu V_\mu +
\hbox{\large$\frac{1}{4}$} \epsilon_{ab} \eta^a \eta^b \Bigr\},
\end{equation*}
which is invariant under the following strictly nilpotent topological 
shift symmetry,
\begin{alignat}{2}
\label{3.2}
&Q^a A_\mu = \psi_\mu^a,
&\qquad
&Q^a E_\mu = 0,
\nonumber
\\
&Q^a V_\mu = - i \psi_\mu^a, 
&\qquad
&Q^a \phi = 0,
\nonumber
\\
&Q^a \zeta^b = 2 \epsilon^{ab} B,
&\qquad
&Q^a \bar{\phi} = 2 \zeta^a,
\nonumber
\\
&Q^a \eta^b = - 2 i \epsilon^{ab} Y +
\hbox{\large$\frac{1}{2}$} \epsilon^{ab} [ \phi, \bar{\phi} ],
&\qquad
&Q^a B = 0,
\nonumber
\\
&Q^a \psi_\mu^b = 2 \epsilon^{ab} E_\mu +
i \epsilon^{ab} \epsilon_{\mu\nu} D^\nu(A - i V) \phi,
&\qquad
&Q^a \bar{B} = [ \eta^a, \phi ],
\nonumber
\\
&Q^a \bar{E}_\mu = i \epsilon_{\mu\nu} D^\nu(A + i V) \zeta^a -
D_\mu(A - i V) \eta^a + i \epsilon_{\mu\nu} [ \psi^{\nu a}, \bar{\phi} ],
&\qquad
&Q^a Y = \hbox{\large$\frac{1}{2}$} i [ \zeta^a, \phi ].
\end{alignat}
Alternatively, since now the underlying topological symmetry is $N_T = 4$,
the action (\ref{3.1}) can be cast also in the following form,
\begin{equation*}
S_T = - \hbox{\large$\frac{1}{2}$} \int d^2x\, {\rm tr} \Bigr\{
i \epsilon^{\mu\nu} B F_{\mu\nu}(A + i V) \Bigr\} + 
\hbox{\large$\frac{1}{2}$} \epsilon_{ab} \,^\star Q^a \,^\star Q^b \bar{X},
\qquad
^\star Q^a = \begin{pmatrix} \,^\star Q \\ \,^\star \bar{Q} \end{pmatrix},
\end{equation*}
with
\begin{equation*}
\bar{X} = - \int d^2x\, {\rm tr} \Bigr\{
\hbox{\large$\frac{1}{2}$} \bar{\phi} \bigr(
B - \hbox{\large$\frac{1}{4}$} i \epsilon^{\mu\nu} 
F_{\mu\nu}(A - i V) \bigr) + i E^\mu V_\mu +
\hbox{\large$\frac{1}{4}$} \epsilon_{ab} \zeta^a \zeta^b \Bigr\},
\end{equation*}
which is invariant under the following strictly nilpotent topological 
co--shift 
symmetry,
\begin{alignat}{2}
\label{3.3}
&^\star Q^a A_\mu = - i \epsilon_{\mu\nu} \psi^{\nu a},
&\qquad
&^\star Q^a \bar{E}_\mu = 0,
\nonumber
\\
&^\star Q^a V_\mu = \epsilon_{\mu\nu} \psi^{\nu a}, 
&\qquad
&^\star Q^a \phi = 0,
\nonumber
\\
&^\star Q^a \eta^b = 2 \epsilon^{ab} \bar{B},
&\qquad
&^\star Q^a \bar{\phi} = 2 \eta^a,
\nonumber
\\
&^\star Q^a \zeta^b = - 2 i \epsilon^{ab} Y -
\hbox{\large$\frac{1}{2}$} \epsilon^{ab} [ \phi, \bar{\phi} ],
&\qquad
&^\star Q^a \bar{B} = 0,
\nonumber
\\
&^\star Q^a \psi_\mu^b = 2 i \epsilon^{ab} \epsilon_{\mu\nu} \bar{E}^\nu +
\epsilon^{ab} D_\mu(A + i V) \phi,
&\qquad
&^\star Q^a B = - [ \zeta^a, \phi ],
\nonumber
\\
&^\star Q^a E_\mu = - D_\mu(A + i V) \zeta^a +
i \epsilon_{\mu\nu} D^\nu(A - i V) \eta^a - 
[ \psi_\mu^b, \bar{\phi} ],
&\qquad
&^\star Q^a Y = - \hbox{\large$\frac{1}{2}$} i [ \eta^a, \phi ].
\end{alignat}
This strongly suggest that both topological symmetries are related to each 
other by a discrete Hodge--type symmetry of the action $S_T$. To elaborate 
this suggestion from (\ref{3.1}) it can be seen that $S_T$ is indeed 
form--invariant under the following duality operation,   
\begin{equation}
\label{3.4}
\star S_T = S_T,
\end{equation}
where the ($\star$) operation is defined by
\begin{equation}
\label{3.5}
\varphi \equiv \begin{bmatrix} 
\partial_\mu & A_\mu & V_\mu & Y
\\
\psi_\mu & \bar{\psi}_\mu & \phi & \bar{\phi} 
\\ 
\zeta & \bar{\zeta} & \eta & \bar{\eta}       
\\
B & \bar{B} & E_\mu & \bar{E}_\mu
\end{bmatrix},
\qquad
\star \varphi = \begin{bmatrix} 
- i \epsilon_{\mu\nu} \partial^\nu & - i \epsilon_{\mu\nu} A^\nu & 
i \epsilon_{\mu\nu} V^\nu & Y
\\
\bar{\psi}_\mu & - \psi_\mu & - \phi & \bar{\phi} 
\\ 
\bar{\eta} & - \eta & \bar{\zeta} & - \zeta       
\\
\bar{B} & B & i \epsilon_{\mu\nu} \bar{E}^\nu & - i \epsilon_{\mu\nu} E^\nu
\end{bmatrix}.
\end{equation} 
By carring out two successive operations of ($\star$) on the generic 
expression $\varphi$ it has the property 
\begin{equation*}
\star (\star \varphi) = \begin{cases} - \varphi & 
\text{for $\varphi = [ \psi_\mu^a, \zeta^a, \eta^a ]$}, \\
\varphi & \text{otherwise}. \end{cases}
\end{equation*}
Furthermore, by inspection of (\ref{3.2}) and (\ref{3.3}) it can be easily 
verified that the topological supercharges ($Q, \,^\star \bar{Q}$) and
($\bar{Q}, \,^\star Q$) are related to each other by the duality transformation
(\ref{3.5}) according to 
\begin{alignat*}{3}
Q ( \varphi) &= \,^\star \bar{Q} \varphi
&\qquad
&\Rightarrow
&\qquad
\,^\star \bar{Q} \varphi &= (\star Q \star) \varphi,
\\
\bar{Q} (\star \varphi) &= - \,^\star Q \varphi
&\qquad
&\Rightarrow
&\qquad
\,^\star Q \varphi &= - (\star \bar{Q} \star) \varphi.
\\
\intertext{Hence, if $S_T$ is invariant under the topological shift 
symmetry (\ref{3.2}), due to the duality invariance (\ref{3.4}), it 
remains 
invariant under the topological co--shift symmetry (\ref{3.3}) as well,
}
Q S_T &= 0 
&\qquad
&\Rightarrow
&\qquad
\,^\star \bar{Q} S_T &= (\star Q \star) S_T = \,^\star Q S_T = 0,
\\
\bar{Q} S_T &= 0 
&\qquad
&\Rightarrow
&\qquad
\,^\star Q S_T &= - (\star \bar{Q} \star) S_T = - \,^\star \bar{Q} S_T = 0.
\end{alignat*}
Therefore, as anticipated by the scheme (\ref{2.1}), the topological 
supercharges ($Q, \bar{Q}$), ($\,^\star Q, \,^\star \bar{Q}$) as well as the 
($\star$) operation can be actually identified with the cohomological 
operators $d$, $\delta$ and the Hodge duality ($\star$) operation, 
respectively. Moreover, by virtue of the anticommutation relations
\begin{equation*}
\{ Q, \,^\star \bar{Q} \} \doteq 2 \delta_G(\phi),
\qquad
\{ \bar{Q}, \,^\star{Q} \} \doteq - 2 \delta_G(\phi)
\qquad
\Rightarrow
\qquad
G \doteq 2 \delta_G(\phi),
\end{equation*}
where the symbol $\doteq$ means that the relations are satisfied only
on--shell, i.e., by taking into account the equations of motions, and where
the gauge transformations $\delta_G(\omega)$ are defined by
$\delta_G(\omega) A_\mu = - D_\mu(A) \omega$ and
$\delta_G(\omega) X = [\omega, X]$, for $X = ( 
V_\mu, \zeta^a, \eta^a, \psi_\mu^a, \phi, \bar{\phi}, B, \bar{B}, 
E_\mu, \bar{E}_\mu, Y )$, 
it follows that on--shell the gauge generator $G$ can be identified 
with the Laplacian $\Delta$. Let us notice, that $\phi$ remains invariant 
under both $Q^a$ and $^\star Q^a$, whereas the vector fields 
$A_\mu - i V_\mu$ and $A_\mu + i V_\mu$ are invariant only under either of
the both supercharges, namely $Q^a$ in the former and $^\star Q^a$ in
the latter case.  

Finally, let us notice that the action (\ref{3.1}) is also invariant
under the following vector and co--vector supersymmetries,
\begin{align}
\label{3.6}
&Q_\mu^a A_\nu = \delta_{\mu\nu} \eta^a - i \epsilon_{\mu\nu} \zeta^a,
\nonumber
\\
&Q_\mu^a V_\nu = - i \delta_{\mu\nu} \eta^a + \epsilon_{\mu\nu} \zeta^a,
\nonumber
\\
&Q_\mu^a \bar{\phi} = 0,
\nonumber
\\
&Q_\mu^a \phi = 2 i \epsilon_{\mu\nu} \psi^{\nu a},
\nonumber
\\
&Q_\mu^a \eta^b = 2 \epsilon^{ab} \bar{E}_\mu -
i \epsilon^{ab} \epsilon_{\mu\nu} D^\nu(A + i V) \bar{\phi},
\nonumber
\\
&Q_\mu^a \zeta^b = - 2 i \epsilon^{ab} \epsilon_{\mu\nu} \bar{E}^\nu - 
\epsilon^{ab} D_\mu(A - i V) \bar{\phi},
\nonumber
\\
&Q_\mu^a \psi_\nu^b = 2 i \epsilon^{ab} \epsilon_{\mu\nu} \bar{B} - 
2 \epsilon^{ab} F_{\mu\nu}(A) -  2 i \epsilon^{ab} D_\mu(A) V_\nu -
2 i \epsilon^{ab} \delta_{\mu\nu} Y +
\hbox{\large$\frac{1}{2}$} \epsilon^{ab} \delta_{\mu\nu} [ \phi, \bar{\phi} ],
\nonumber
\\
&Q_\mu^a \bar{B} = i \epsilon_{\mu\nu} D^\nu(A + i V) \eta^a,
\nonumber
\\
&Q_\mu^a B = 2 D_\mu(A) \zeta^a -
i \epsilon_{\mu\nu} D^\nu(A - i V) \eta^a + [ \psi_\mu^a, \bar{\phi} ],
\nonumber
\\
&Q_\mu^a \bar{E}_\nu = 0,
\nonumber
\\
&Q_\mu^a E_\nu = D_\mu(A + i V) \psi_\nu^a - D_\nu(A + i V) \psi_\mu^a +
\delta_{\mu\nu} D_\rho(A - i V) \psi_\rho^a -
[ \delta_{\mu\nu} \zeta^a + i \epsilon_{\mu\nu} \eta^a, \phi ],
\nonumber
\\
&Q_\mu^a Y = i D_\mu(A - i V) \eta^a +
\hbox{\large$\frac{1}{2}$} \epsilon_{\mu\nu} [ \psi^{\nu a}, \bar{\phi} ]
\\
\intertext{and}
\label{3.7}
&^\star \bar{Q}_\mu^a A_\nu = i \epsilon_{\mu\nu} \zeta^a - 
\delta_{\mu\nu} \eta^a,
\nonumber
\\
&^\star \bar{Q}_\mu^a V_\nu = - \epsilon_{\mu\nu} \zeta^a + 
i \delta_{\mu\nu} \eta^a,
\nonumber
\\
&^\star \bar{Q}_\mu^a \bar{\phi} = 0,
\nonumber
\\
&^\star \bar{Q}_\mu^a \phi = - 2 i \epsilon_{\mu\nu} \psi^{\nu a},
\nonumber
\\
&^\star \bar{Q}_\mu^a \zeta^b = - 2 i \epsilon^{ab} \epsilon_{\mu\nu} E^\nu +
\epsilon^{ab} D_\mu(A - i V) \bar{\phi},
\nonumber
\\
&^\star \bar{Q}_\mu^a \eta^b = 2 \epsilon^{ab} E_\mu + 
i \epsilon^{ab} \epsilon_{\mu\nu} D^\nu(A + i V) \bar{\phi},
\nonumber
\\
&^\star \bar{Q}_\mu^a \psi_\nu^b = 2 i \epsilon^{ab} \epsilon_{\mu\nu} B + 
2 \epsilon^{ab} F_{\mu\nu}(A) -  
2 i \epsilon^{ab} \epsilon_{\mu\rho} \epsilon_{\nu\sigma} D^\rho(A) V^\sigma -
2 i \epsilon^{ab} \delta_{\mu\nu} Y -
\hbox{\large$\frac{1}{2}$} \epsilon^{ab} \delta_{\mu\nu} [ \phi, \bar{\phi} ],
\nonumber
\\
&^\star \bar{Q}_\mu^a B = - D_\mu(A - i V) \zeta^a,
\nonumber
\\
&^\star \bar{Q}_\mu^a \bar{B} = - 2 i \epsilon_{\mu\nu} D^\nu(A) \eta^a +
D_\mu(A + i V) \eta^a + [ \psi_\mu^a, \bar{\phi} ],
\nonumber
\\
&^\star \bar{Q}_\mu^a E_\nu = 0,
\nonumber
\\
&^\star \bar{Q}_\mu^a \bar{E}_\nu = D_\mu(A - i V) \psi_\nu^a - 
D_\nu(A - i V) \psi_\mu^a +
\delta_{\mu\nu} D_\rho(A + i V) \psi_\rho^a -
[ \delta_{\mu\nu} \zeta^a + i \epsilon_{\mu\nu} \eta^a, \phi ],
\nonumber
\\
&^\star \bar{Q}_\mu^a Y = \epsilon_{\mu\nu} D^\nu(A + i V) \zeta^a +
\hbox{\large$\frac{1}{2}$} \epsilon_{\mu\nu} [ \psi^{\nu a}, \bar{\phi} ],
\end{align}
respectively. These symmetries, together with the topological shift and 
co--shift symmetries (\ref{3.2}) and (\ref{3.3}), obey the topological 
superalgebra 
\begin{alignat*}{2}
\{ Q^a, Q^b \} &= 0,
&\qquad
\{ Q_\mu^a, Q_\nu^b \} &\doteq 
2 i \epsilon^{ab} \epsilon_{\mu\nu} \delta_G(\bar{\phi}),
\\
\{ Q^a, \,^\star Q^b \} &\doteq 
2 \epsilon^{ab} \delta_G(\phi),
&\qquad
\{ Q_\mu^a, \,^\star Q_\nu^b \} &\doteq 
- 2 i \epsilon^{ab} \epsilon_{\mu\nu} \delta_G(\bar{\phi}),
\\
\{ \,^\star Q^a, \,^\star Q^b \} &= 0,
&\qquad
\{ \,^\star Q_\mu^a, \,^\star Q_\nu^b \} &\doteq 
2 i \epsilon^{ab} \epsilon_{\mu\nu} \delta_G(\bar{\phi}),
\\
\{ Q^a, Q_\mu^b \} &\doteq 2 \epsilon^{ab} (
\partial_\mu + \delta_G(A_\mu - i V_\mu)),
&\qquad
\{ \,^\star Q^a, Q_\mu^b \} &\doteq 2 i \epsilon^{ab} 
\epsilon_{\mu\nu} ( \partial^\nu + \delta_G(A^\nu + i V^\nu)),
\\
\{ Q^a, \,^\star Q_\mu^b \} &\doteq - 2 \epsilon^{ab} (
\partial_\mu + \delta_G(A_\mu - i V_\mu)),
&\qquad
\{ \,^\star Q^a, \,^\star Q_\mu^b \} &\doteq - 2 i \epsilon^{ab} 
\epsilon_{\mu\nu} ( \partial^\nu + \delta_G(A^\nu + i V^\nu)).
\end{alignat*}
Thereby, the vector supercharges $Q_\mu^a$ and $^\star Q_\mu^a$ are related 
to each other by the duality operation (\ref{3.5}) according to
\begin{equation*}
Q_\mu^a = \begin{pmatrix} Q_\mu \\ \bar{Q}_\mu \end{pmatrix},
\qquad
^\star Q_\mu^a = \begin{pmatrix} ^\star Q_\mu \\ ^\star \bar{Q}_\mu 
\end{pmatrix},
\qquad
\hbox{with}
\qquad
^\star \bar{Q}_\mu = \star Q_\mu \star,
\qquad
^\star Q_\mu = - \star \bar{Q}_\mu \star.
\end{equation*}

It is worth mentioning that the form of the action (\ref{3.1}) is completely 
specified by the topological shift symmetries $(Q^a, \,^\star Q^a)$ and the 
vector supersymmetries $(Q_\mu^a, \,^\star \bar{Q}_\mu^a)$, fixing all the 
relative numerical coefficients in (\ref{3.1}) and, therefore, allowing, 
in particular, for a single coupling constant.

In conclusion we remark that in a similar spirit 
one can study another topological twist of $N = 16$, $D = 2$ SYM, i.e., 
the dimensional reduction of $N = 1$, $D = 10$ SYM to $D = 2$, with an 
underlying $N_T = 8$ equivariant cohomology, respectively. 
It describes a topological gauge theory associated to D1--branes on 
holomorphic curves in K3s \cite{16}. By restricting to the Euclidean 
space--time this topological model provides other, but more 
involved example of a Hodge--type cohomological theory. The same model
can also be obtained by a dimensional reduction of the $N_T = 4$
equivariant extension of the Blau--Thompson model \cite{19} to $D = 2$. 

%%%%%%%%%%%%%%%%%%%%%%%%%%%%%%%%%%%%%%%%%%%%%%%%%%%%%%%%%%%%%%%%%%%%%%%%%%%%%

\end{document}